\documentclass[twocolumn,prl,amsmath,amssymb,superscriptaddress,showpacs]{revtex4-1}
\pdfoutput=1

\usepackage{amsmath, amsthm, amssymb}
\usepackage{graphicx}
\usepackage{dcolumn}
\usepackage{bm}
\usepackage{color}

\def\textbf#1{{\bf #1}}
\def\be{\begin{equation}}
\def\ee{\end{equation}}
\def\ben{\begin{eqnarray}}
\def\een{\end{eqnarray}}
\def\eea{\end{array}}
\def\bea{\begin{array}}

\newcommand{\bei}{\begin{itemize}}
\newcommand{\eei}{\end{itemize}}

\def\ra{\rangle}
\def\la{\langle}

\begin{document}

\newcommand{\eg}{{\it{e.g.~}}}
\newcommand{\ie}{{\it{i.e.~}}}
\newcommand{\etal}{{\it{et al.}}}
\newcommand{\daniel}[1]{{\color{red} #1}}
\newcommand{\leandro}[1]{{\color{blue} #1}}
\newcommand{\marco}[1]{{\color{green} #1}}


\newcommand{\ci}[2]{{I(#1\rangle #2)}}
\newcommand{\disc}[2]{{D(#1 | #2)}}
\newcommand{\discinf}[2]{{D^\infty(#1 | #2)}}
\newcommand{\dc}[2]{\chi_{\text{DC}}(#1\rangle #2)}
\newcommand{\Ddc}[2]{\Delta_{\text{DC}}(#1\rangle#2)}
\newcommand{\tec}[2]{\Gamma(#1\rangle#2)}
\newcommand{\tecinf}[2]{\Gamma^\infty(#1\rangle#2)}
\newcommand{\eof}[2]{E_F(#1:#2)}


\author{D. Cavalcanti}
\email{dcavalcanti@nus.edu.sg}
\affiliation{Centre for Quantum Technologies, National University of Singapore, 2 Science Drive 3, Singapore 117542}
\author{V. Scarani}
\affiliation{Centre for Quantum Technologies, National University of Singapore, 2 Science Drive 3, Singapore 117542}
\affiliation{Department of Physics, National University of Singapore, 2 Science Drive 3, Singapore 117542}

\title{A note on Phys. Rev. Lett. 105, 170404 (2010)}
\begin{abstract}

\end{abstract}

\maketitle

In \cite{paper}, two Bell inequalities suitable for nonlocality tests with continuous variables were proposed. In order to test these Bell inequalities, two parties, A and B, measure 3 observables each, $X_j$, $Y_j$, or $N_j (j=A, B)$, and the inequalities read:\be\label{In1}
(\la X_A X_B\ra+\la Y_A Y_B \ra)^2+(\la X_A Y_B\ra-\la Y_A X_B\ra )^2\leq \la N_A N_B\ra
\ee
\be\label{In2}
(\la X_A X_B\ra+\la Y_A Y_B \ra)^2+(\la X_A Y_B\ra-\la Y_A X_B\ra )^2\leq \la N_A\ra\la N_B\ra.
\ee In the case of modes of the electromagnetic field, $X_j$ and $Y_j$ correspond to two orthogonal quadratures of the electromagnetic field, and $N_j=a^\dagger_j a_j$ to the number operator, being $a_j$ the annihilation operator of the mode $j$.

However, the inequalities \eqref{In1} and \eqref{In2} are not Bell's inequalities in the usual sense: there are local classical models that violate these inequalities. An example of such a model is a classical source assigning the following values to the observables:
\ben X_A=X_B=Y_A=Y_B=1&,&N_A=N_B=0\,.\label{lhv}\een This proves that assumptions on the physical system (or on the class of local variable models to be excluded) are made in the derivation of the inequalities.

Indeed, the authors of \cite{paper} acknowledge such assumptions in their footnote [22], when they state that classical fields should satisfy
\ben
X_j^2+Y_j^2=N_j.\label{cond}
\een This deserves three comments:
\begin{enumerate}
\item If the constraints (\ref{cond}) are actually inserted in the inequality instead of the $N$, the resulting expressions cannot be violated by any state \cite{salles}. The authors are aware of this, and this is why they chose to consider $N$ as an independent measurement. But then, those constraints are additional criteria, that the inequalities themselves do not capture.
\item At the abstract level of all possible local variable models, therefore, one can only claim that the inequalities exclude those models that satisfy the constraints (\ref{cond}). A careful study may lead to less strict requirements; but certainly, the inequalities \eqref{In1} and \eqref{In2} will never exclude \textit{all} possible variable models, given the explicit counterexample (\ref{lhv}).
\item Let us now assume that the measured physical system is indeed the electromagnetic field. A classical field satisfies (\ref{cond}) only if \textit{exactly the same modes} are probed when one measures $X$, $Y$ and $N$. Again, one can possibly weaken the assumptions. But there is a classical optical implementation that leads to the local variable model (\ref{lhv}): suppose that $X$ and $Y$ are measured for horizontally polarized light and $N$ for vertically polarized light; then, one can prepare a horizontally polarized field that gives $X_A=X_B=Y_A=Y_B=1$, which certainly will produce $N_A=N_B=0$. 
\end{enumerate}

In summary, the inequalities \eqref{In1} and \eqref{In2} proposed in \cite{paper} rule out only a restricted class of local variable models; or, equivalently, their violation demonstrates nonlocality only under assumptions about the physical implementation \cite{footnote}. In particular, they cannot be used as a device-independent test: a malicious adversary may engineer a fake violation with classical means, as demonstrated here.

Finallly, from a positive point of view, the ideas presented in \cite{paper} might be useful in different contexts other that nonlocality tests. For instance, the fact that the equality \eqref{cond} can be violated in quantum mechanics suggests it as a test of quantumness. Indeed, this idea could be used to obtain quantitative estimates of $\la [X,Y]\ra$ (or $\la [a^\dagger,a]\ra$ ), in a similar spirit as in Refs. \cite{Bellini}.

\begin{acknowledgements}
We thank the National Research Foundation and the Ministry of Education of Singapore.\end{acknowledgements}


\begin{thebibliography}{10}
\bibitem{footnote}In our opinion, if it comes to trusting models of the degrees of freedom under study, one might just as well trust the model and experimental characterization of the detection process. This would show that is no correlation between the firing of a detector and the measurement that is chosen, making of the detection loophole an artificial issue.
\end{thebibliography}
\end{document}